\documentclass[
    ,final            
  ]
  {aipproc}

\layoutstyle{6x9}

\usepackage{graphicx}
\usepackage{amsfonts}
\usepackage{amssymb}
\usepackage{amsmath}
\usepackage{epsfig}

\newcommand{\be}{\begin{equation}}
\newcommand{\ee}{\end{equation}}
\newcommand{\bea}{\begin{eqnarray}}
\newcommand{\eea}{\end{eqnarray}}
\newcommand{\4}{{\,4}}
\newcommand{\2}{{\,2}}

\begin{document}

\title{Anomaly Mediated Supersymmetry Breaking Demysitified
 \footnote{ Based on the talk given by D.~W.~Jung at the SUSY09, Boston, USA and
 the work \cite{Jung:2009dg}  by the same authors.} }


\author{Dong-Won Jung}{
  address={Physics Department and CMTP, National Central University,
Jhongli, 32054, Taiwan.}
}

\author{Jae Yong Lee}{
  address={School of Physics, KIAS,
Dongdaemun-Gu, Seoul 130-722, Korea.}
}

\begin{abstract}

We reinterpret anomaly-mediated supersymmetry breaking from a field-theoretic perspective
in which superconformal anomalies couple to either the chiral compensator or the $U(1)_R$ vector superfield.
As supersymmetry in the hidden sector is spontaneously broken by non-vanishing vacuum expectation values
of the chiral compensator F-term and/or the $U(1)_R$ vector superfield D-term,
the soft breakdown of supersymmetry emerges in the visible sector.
This approach is physically more understandable compared with the conventional approach
where the chiral compensator is treated on the same footing as a spurion in gauge-mediated
supersymmetry breaking scenario.

\end{abstract}

\maketitle


\section{Introduction}

 Anomaly Mediated Supersymmetry Breaking (AMSB) scenario~\cite{Randall:1998uk,Giudice:1998xp} is 
 fascinating among the supersymmetry (SUSY) breaking models. In that scenario,
{\it the chiral compensator} plays a central role not only
as the mediator of SUSY breaking but also as the superfield spurion
whose auxiliary component sets the SUSY breaking mass scale.

 In this work, we interpret the chiral compensator as a supersymmetric extension
of the Nambu-Goldstone boson (NGB) associated with broken superconformal symmetry.
 This leads to the coupling of the chiral compensator to the chiral anomaly
supermultiplet (CASM)~\cite{Ferrara:1974pz} through the quantum effects.
When the F-term vacuum expectation value (VEV) of the chiral compensator is turned on
by the SUSY breaking in the hidden sector~\cite{Pomarol:1999ie},
the soft breaking of SUSY in the visible sector emerges.

 As the CASM couples to the chiral compensator
the linear anomaly supermultiplet (LASM)~\cite{Sohnius:1981tp,Akulov:1976ck} also
couples to a superfield which is referred to as {\it the $U(1)_R$ vector superfield}.
In a similar manner to the chiral compensator, the $U(1)_R$ vector superfield forms a genuine vector superfield.
When the D-term of the $U(1)_R$ vector superfield gets a non-vanishing VEV due to
the hidden sector~\cite{Pomarol:1999ie}, the soft terms in the visible sector take place as well.

\section{Chiral anomaly supermultiplet}

 Let's start with the CASM.
%
 Suppose that supercurrents are anomalous, {\it i.e.},
\be\label{eq:anom.set}
\xi_\alpha\equiv\gamma_m S^m_\alpha\neq 0,\quad
\mathring{t}\equiv T^m_m\neq 0,\quad
\mathring{r}\equiv\partial^m j^R_m\neq 0, \ee
where  $T_{mn}$ is energy momentum tensor,  $S^m_\alpha$ is SUSY current, and 
 $j^R_m$ is $R$-current respectively.
It is known that they form the CASM with auxiliary fields $a$ and $b$~\cite{Clark:1978jx}, 
\be\label{eq:anom.sup}
{\cal X}(x,\theta)\equiv {\cal A}(x)+\sqrt{2}\theta\xi(x)+\theta\theta{\cal F}(x),
\ee
where ${\cal A}=a+ib$ and ${\cal F}=\mathring{t}+i\mathring{r}$

 The next question is what kind of superfield couple to the CASM. The answer is as follows:
 the trace anomaly $\mathring{t}$ couples to dilaton, $\varrho(x)=1/2\ln\mbox{det}[e^m_a]$,  
 the $U(1)_R$ anomaly $\mathring{r}$ to the local $R$-symmetry, $\delta(x)$ (NGB of $U(1)_R$), and 
 the SUSY anomaly $\xi_\alpha$ to  the dilatino ($ \sim \bar\Psi_\alpha(x) =\sigma_m^{\alpha\dot\alpha}\bar\psi^m_{\dot\alpha}(x)$).
We come to conclude that the chiral compensator couples to the CASM ;
\be\label{eq:full.comp}
\chi^3(x,\theta) \equiv  e^{2\varrho(x)+2i\delta(x)}
[1+\sqrt{2}\theta \bar\Psi(x)+\theta\theta M^\ast(x)],\ee
In reality, the chiral compensator is invrinat measure,
\be d^\4 x'd^\2\theta' \chi'^3(x',\theta')=d^\4 x d^\2\theta\chi^3(x,\theta). \ee
  We can comprehend this property by decomposing the chiral compensator as
\be
d^\4 xd^\2\theta\chi^3(x,\theta)=
\bigg\{d^\4 x\,e^{2\varrho(x)}\bigg\}\bigg\{d^\2\theta\,e^{2i\delta(x)}[1
+\sqrt{2}\theta^\alpha \bar\Psi_\alpha(x)+\theta\theta M^\ast(x)]\bigg\}.
\ee
Finally, the trace anomaly with invaraint measure, the action is given in explicit form
\be
S_{\cal X}=\int d^\4 x\,d^\2\theta\,\chi^3(x,\theta) {\cal X}(x,\theta)+ c.c.
\ee

 Non-vanishing VEV $\langle \chi^3 \rangle=1+\theta\theta \langle M\rangle$
 leads to soft terms in the visible sector.
To see this more clearly, we can rewrite the action with component fields,
\be\label{eq:anom.act}
S_{\cal X}=\int d^\4 x\, [e^{2\varrho+2i\delta}(M^\ast{\cal A}+\bar\Psi\xi+{\cal F})+c.c.],
\ee
where the first term with nonzero-VEV $M$ leads soft term and nonzero
 $\langle \delta \rangle $ will give a term with $ F\widetilde{F}$, which describes the axion.

\section{Linear anomaly supermultiplet}

 Assume that
\be \partial^m j^R_m=0\label{eq:r.curr}, \quad 
\xi_\alpha=\gamma_m S^m_\alpha\neq 0,\quad \mathring{t}=T^m_m\neq 0. \ee
 In this case, $ j^R_m,~ \xi_\alpha,$ and  $\mathring{t}$ form the LASM~\cite{Sohnius:1981tp,Akulov:1976ck}.
 Most generally, the LASM is written
\bea
L(x,\theta,\bar\theta)&=&C(x)+i\theta\Xi-i\bar\theta\bar\Xi
+\theta\sigma^m\bar\theta j^R_m(x)\nonumber\\
&&-\frac{1}{2}\,\theta\theta\bar\theta\bar\sigma^m\partial_m\Xi(x)
-\frac{1}{2}\,\bar\theta\bar\theta\theta\sigma^m\partial_m\bar\Xi(x)
-\frac{1}{4}\,\theta\theta\bar\theta\bar\theta\, \Box C(x).
\eea
Here we can identify 
 $\mathring{t} \sim \Box C$, 
$\xi_\alpha \sim \sigma^m_{\alpha\dot\alpha}\partial_m\bar\Xi^{\dot\alpha}$.

 Let's think about what superfield it couples to.
Using the following supersymmetric generalization of $U(1)_R$ gauge transformations:
\be\label{eq:u1group} {\cal V}\to{\cal V}+\Lambda+\Lambda^+, \ee
with $\Lambda$ being a chiral field $(\bar D \Lambda=0)$.
The most generally,
\bea
{\cal V}(x,\theta,\bar\theta)&=&s(x)+i\theta \omega(x)-i\bar\theta\bar\omega(x)
+\frac{i}{2}\,\theta\theta[p(x)+iq(x)]-\frac{i}{2}\,\bar\theta\bar\theta[p(x)-iq(x)]\nonumber\\
&&-\theta\sigma^m\bar\theta \nu_m(x)+i\theta\theta\bar\theta[\bar\tau(x)
+\frac{i}{2}\,\bar\sigma^m\partial_m\omega(x)]\nonumber\\
&&-i\bar\theta\bar\theta\theta[\tau(x)+\frac{i}{2}\,\sigma^m\partial_m\bar\omega(x)]
+\frac{1}{2}\,\theta\theta\bar\theta\bar\theta[{\mathbf d}(x)+\frac{1}{2}\,\Box s(x)].
\eea
In the Wess-Zumino gauge where $s=\omega=p=q=0$, and we identify 
$\nu_m$, $\tau_\alpha$ and ${\mathbf d}$ are the gauge field, gaugino, and $D$-term,
respectively, for $U(1)_R$.  Note that ${\mathbf d}$ and $\tau_\alpha$ are gauge invariant component fields in ${\cal V}$.
 For these fields, we can roughly identify
\bea  \nu_{\alpha\dot\alpha} &=& \Psi_\alpha\bar\Psi_{\dot\alpha}, \quad 
\tau_\alpha = \bar\Psi_\alpha M, \quad  
{\mathbf d} = M^\ast M, \eea
where $\nu_{\alpha\dot\alpha}=\nu_m \sigma^m_{\alpha\dot\alpha}$.

 Finally, the gauge invariant and supersymmetric action is
\be\label{eq:super.ac2} S_L=\int d^\4 xd^\2\theta d^\2\bar\theta\, {\cal V}L,\ee
 which is rewritten with component fields,
\be \label{eq:lin.coup}
S_L=\int d^\4 x\,(\nu^m j^R_m +\frac{1}{2}\,{\mathbf d}\,C+\tau\,\Xi+\bar\tau\,\bar\Xi). \ee
 Here the second term is only relevant for the soft sfermion masses. It is the footprints of
conformal anomaly.

\section{THE MSSM soft terms}
 We can derive the soft terms of the MSSM with the formulas given in the previous sections. 
Let's assume that $\langle M^* \rangle = m_{3/2}$ and $ {\mathbf d} = m_{3/2}^2$.
First we consider the CASM from the gauge kinetic term.
\bea
{\cal X} = \frac{\beta(g)}{2g}\,W^{a\alpha} W^a_\alpha,
\eea
where $\beta (g)$ is the MSSM $\beta$-function.
Then from the Eq. (\ref{eq:anom.act}), the gaugino mass terms are derived as
\be
M_\lambda = \frac{\beta(g)}{g} m_{3/2}.
\ee

  Secondly, the CASM from the Yukawa terms reads \\
\be
{\cal X} = \frac{1}{3!}(\gamma_{i} +\gamma_j +\gamma_k ) y^{ijk} \phi_{i} \phi_j \phi_k.
\ee
Again from the Eq. (\ref{eq:anom.act}) trilinear A term is generated as
\bea
A_{ijk} = -(\gamma_i +\gamma_j+\gamma_k) y^{ijk} m_{3/2}.
\eea
we also can supersymmetrize the dilaton with which we can derive the dilaton(dilatino) and 
the CASM interactions.

 Finally, let's consider the LASM.It is evaluated at two loop,
\bea L = -\frac{1}{4}\dot\gamma_i\,{\phi_i}^{\!+}\phi_i, \quad {\rm with } \quad 
\dot\gamma_i =  \frac{\partial \gamma^j_i}{\partial \ln \mu}.
\eea
Similarly from the Eq. ({eq:lin.coup}), the soft mass terms read 
\bea
m_i^2 = \frac{1}{4} \dot{\gamma^i_i} m^2_{3/2}.
\eea

\section{Summary}

 We present the novel field-theoretical understanding of the
Anomaly Mediated Supersymmetry breaking scenario. It is more understandable compared with the conventional spurion
method.  We can reproduce the results with the anomaly superfields interactions.
In addition to that, the dilaton-interaction action is acquired, which can be used for study of hidden sector.


\begin{theacknowledgments}
\end{theacknowledgments}

\end{document}